# Merging Object and Process Diagrams for Business Information Modeling

Patrick Chénais[1]

[1] formerly worked for IT Services at the University of Bern, Switzerland,
now working at the Federal Office of Information Technology, Systems and
Telecommunication in Bern FOITT, Switzerland

**Abstract:** While developing an information system for the University of Bern, we were faced with two major issues: managing software changes and adapting Business Information Models. Software techniques well-suited to software development teams exist, yet the models obtained are often too complex for the business user. We will first highlight the conceptual problems encountered while designing the Business Information Model. We will then propose merging class diagrams and business process modeling to achieve a necessary transparency. We will finally present a modeling tool we developed which, using pilot case studies, helps to show some of the advantages of a dual model approach.

## 1   Presentation of the paper

After a short presentation in chapter 2 of our objectives and the problems we encountered, we shall examine in chapter 3 the general task of users in an administrative environment. We will here point out the importance of data transformation that is not adequately conveyed by static class diagrams. Chapter 5 proposes a method and a tool for merging static and dynamic aspects of the modeling of business processes, with the aim of dealing with both aspects in one overview diagram that is more accessible to non-IT experts. The method is based on previous work on the QOBJ paradigm, which is introduced in chapter 4. Chapter 6 discusses the method and proposes a rigorous process that allows the integration of object and process modeling. Chapter 7 develops a case study applying the method.

## 2   Introduction

As in any large company or administration, the administrative tasks of the University of Bern are distributed between various administrative services. Each service has a dedicated information processing system responsible for managing its data. The dedicated computer systems have two functions: 1) to collate the data, and 2) to provide tools for monitoring, modifying and processing the data. To fulfill these functions, each service uses data collated by other services. A major issue for the development of administrative applications is the exchange of data between dedicated computer systems. Our objective is to develop an information system on which all of the processing systems can be based. This information system will reduce the complexity of interfaces, simplifying integration and maintenance.

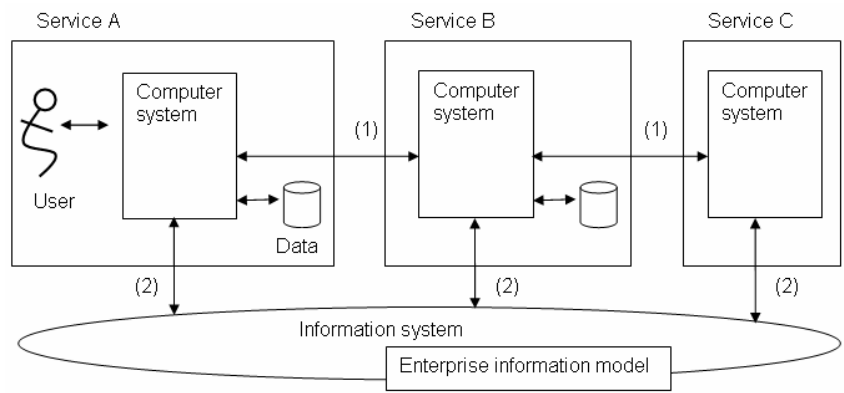

**Fig. 1. Computer systems and information system** - A computer system is a piece of hardware with which users interact. Computer systems exchange data by the way of a point to point (1) communication approach. In order to reduce the complexity of the exchanges between services, our objective is to develop an information system that acts as the provider of information (2). This information system consists of a hardware infrastructure, basically an Enterprise Service Bus, and an Enterprise Information Model, which is a logical description of the data for the whole enterprise.

In a point to point communication between systems, it is easy to interpret the semantic contents of data, as the source of the information is clearly identified. This is no longer the case when an information system provides the data, because it may aggregate information coming from different sources. A necessary step is to design an enterprise information model.

It was difficult to establish such a model for the university because we could not find an accurate definition of the terms used. We also observed that an object model even in the very simple form of a UML Class Diagram was too technical for users. (We define a user as any person in charge of some business aspects of the university. Thus they are operators of the computer systems, but also experts in administrative and academic processes of the university.) The information model is either well-suited nor, in the end, useful to a description of the user's tasks. In other terms, the model was inappropriate because users did not recognize their tasks or, consequently, how the model could be useful to them. The use of the model as a common dictionary and database did not register with the users. They considered the object model a specific tool for computer specialists. The first very simple, imperfect and incomplete object model had to undergo major transformations. Still at the beginning of the project, we had to consider improvements to the object model and its software implementation [1] [2] [3] [4]. The next task was to ensure the acquisition of the user's business knowledge. So we faced a double issue: 1) how to master the software changes, and 2) how to establish a dialogue with users in order to acquire their business knowledge? We will now focus on the second point.

## 3 Business information model

Why did users not find it easier to understand the object model? To answer this question, we must examine the nature of the administrative tasks and the users' activities.

It is difficult to give a common definition of the administrative activities as they can vary so much. The concept of a file however is common to them all: budgetary and financial files, or, for the university, student or exam files. Setting up files with all the necessary information for decisional authorities constitutes the main activity of administrative services. Employees (users of computer systems) manage and process information which moves between several services and hierarchical levels. At the company level this information has a permanent character and a long lifespan. For instance, information related to one person within a customer relations management system (mailing, feedback, purchases, requests for information, calls to the after-sales service, contact details, etc.) is entered by users belonging to various different services. The data structure is permanently transformed, but the data never disappears from the system.

A simple example will help us to illustrate the phenomena. Let's consider the organization of a meeting. The process of organizing a meeting is shown below.

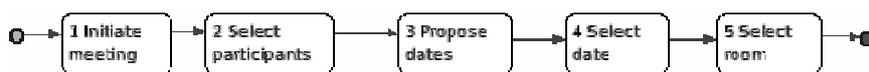

**Fig. 2. Meeting organization process** - The organization of a meeting follows several stages. After a phase of initialisation when name and subject of the meeting are defined, the meeting organizer selects the participants (2) and proposes dates (3). Participants select dates to their personal convenience (3). The meeting organizer selects one of the proposed dates (4) and finally reserves a room (5).

At the second stage (*2 Select participants*) the meeting organizer's task is to set up a list of persons. According to the practices of simulation queuing networks, an object *Meeting* crosses the process. This object (token, a token being an object that moves through processes) follows the progress of the process of creating a meeting. At the second stage a list of persons is associated with that token. Starting with a given list of persons (participants at the meeting), the task at the third stage (*3 Propose date*) is to propose dates and collect the possible choices of participants. Associated with the token are now: a list of participants and a list of dates. By this point, relations will have been established between participants and dates. So, it is evident that the structures of the data related to the organization of the meeting vary as the process goes on. As further illustration, consider for instance the information concerning the location which will be added in the fifth stage (*5 Select room*).

We could generally say that the transformation or completion of data, which is the objective of a process, may require intermediate data (and data structures) to be performed. Users are interested only in data related to their own activity. If we want the data model to be understandable and valuable to their tasks, it becomes necessary to show only the data with which they are concerned. The issue now is to find an appropriate way to do this.

A first solution would be to associate a dedicated UML Class Diagram to each activity. This though involves major drawbacks. Firstly, we arrive at not a single Enterprise Information Model, but as many models as there are activities or different tasks in the company. Secondly, how are the links between the data models and the process models managed and saved? Another system is required to maintain these relations. How is this third entity modeled?

We can draw the following conclusions:
- The business data structures have a long lifespan at the company level and (therefore) flow through many transformation processes. The activity of an employee is a part of the process. It consists of modifying the data structure but covers only a very limited domain.
- The data structures must be defined in relation to the user's activity so that he can easily understand them. Only the data relevant to the user's activity should be shown.

The first conclusion confirms the merit of the proven technique of Structured Analysis (SA) ([22]), which uses Data Flow Diagrams (DFD), and places data transformations at the heart of the method. However, data consistency is poorly accounted for by this technique, firstly because the data dictionary is primarily a means for identifying refinement of the data flows, and secondly because the data store has no formal purpose other than to contain data. Therefore object-oriented methods are likely to bring an interesting complement to the structured analysis and, more generally, to the process-oriented methods.

The second conclusion points to working out the data model in conjunction with the process model. A close relationship (bond) must be established between object models and process models. For instance, UML Class Diagrams and UML Activity Diagrams (or BPMN Diagrams which are formally equivalent) should be merged into one Business Information Model.

## 4   Integrating Object-oriented methods and Process modeling

The issue of the integration of object-oriented (OO) methods and process modeling approaches arose in the 1980's when OO techniques began to be widely used. We do not wish to draw up a complete overview of these attempts, but in order to identify the possible benefit of the object-oriented methods in processing modeling techniques, we outline two main strategies.

Starting from a DFD, the first strategy consists of transforming the data containers (i.e. Data Store) into objects. The processes of the Data Flows then determine the methods of the objects ([11]). This approach was extended to distributed systems, with technologies overcoming the barrier of networks (CORBA, Microsoft DNA with DCOM, etc.). This approach does not provide any solution to the issue of data transparency, because the process approach completely disappears.

The second strategy is widely applied in simulation networks (event driven simulation). It consists of using objects for data exchange between distributed processes. Web Services (and Service Oriented Architecture SOA) implement a similar strategy: objects are exchanged between services using XML structures. This is primarily the approach we adopted in the 1990's in order to accommodate a static object model and a network-oriented dynamic model in the same system (an elevator control system). We developed a unifying concept called QOBJ (queue + object) ([8] [9] [10]).

Curiously, the effort of integrating the object-oriented and process approaches was not further continued. Instead, object-oriented techniques have developed, overcoming the concurrent approach which lead to the abandoning of Structured Analysis. With the strong development of Workflow Management and Business Process Modeling (BPM) in the 1990s, the interest in process approach has reappeared. It became essential to introduce Activity Diagrams to the UML language. Although some people feel that using activity diagrams is not object-oriented [14] and therefore activity diagrams constitute a foreign body within an object-oriented method, their introduction was justified by their usefulness.

Which method is most appropriate when it comes to the integration of the object-oriented method and process modeling? Although it does not constitute a well-known standard, we have chosen the QOBJ paradigm. The reasons are: 1/ we have a very good experience with this paradigm. It has been successfully applied within projects in the field of high-rise elevator control systems. One major innovative project of the Schinder company during the nineties was the project (and product) Artificial Intelligence Traffic Processor AITP$^{TM}$ [6] [23], which included a Virtual Elevator System [7] developed with the QOBJ approach: and, 2/ QOBJ paradigm allows the co-existence of three approaches: object modeling, process modeling and multi agent (distributed) control. Hence a major step towards merging the approaches was made.

One key factor has been to recognize that the issues encountered during the modeling of administrative processes and their monitoring were similar to those we faced while developing controls for elevator group systems. Since we solved these issues applying the QOBJ paradigm, it followed to use the same solution for the needs of Business Process Modeling.

QOBJs have the necessary consistency that Structured Analysis lacks. One remaining drawback to this approach is inherent to the object-oriented approach: the difficulty of modifying an object's pre-defined structures. As we have seen with the example of the meeting organization process (figure 2), this is necessary because users should only have access to the data relevant to their own activities. We developed a modeling tool to resolve this issue and will describe this tool in the next chapter.

## 5  Topologos: Merging data and process model

Topologos is a modeling tool which enables the integration of object and process models. The software is written in Squeak, a multimedia open source (and free) dialect of Smalltalk [12]. Models are saved using XML files. The integrated graphical user interface of Topologos allows for a faithful representation of a model's contents. We present here this user interface. A Web interface (developed with Seaside) is also available. A strong separation of models and their graphical representations allows models to be manipulated without a user interface. Application programming interfaces (API) can also easily be developed.

Topologos is based on event-driven simulation techniques and object flow diagrams. The QOBJ approach serves as the basis for Topologos. According to the approach, UML Activity Diagrams, for instance, are QOBJ networks. Some adaptations of the QOBJ paradigm were necessary in order to realize Topologos. (The main principles of QOBJ modeling are outlined below.)

**Table 1. The QOBJ Paradigm**

> A QOBJ is an object which has a queue (named place) containing other objects (i.e. QOBJs) crossing the network. No differentiation is made between static and mobile objects, nor between active and passive objects. All are QOBJs. A QOBJ may have a pilot, another QOBJ responsible for the processing of QOBJs entering its queue. A Service (i.e. a list of instructions) is allocated to each Pilot. Pilots and their Services allow for a simple implementation of distributed processing. The QOBJ run-time environment consists of a communication Kernel based upon event driven simulation techniques and running in real-time or simulated-time. With this Kernel system dynamics can be simulated, and real-time systems can be monitored and controlled without any modifications to the application software. Callback mechanisms are used to inform applications about model dynamic changes.

The first adaptation is to divide a QOBJ into two parts. The first part, called a Node, constitutes, as the QOBJ does, the nodes of the network.

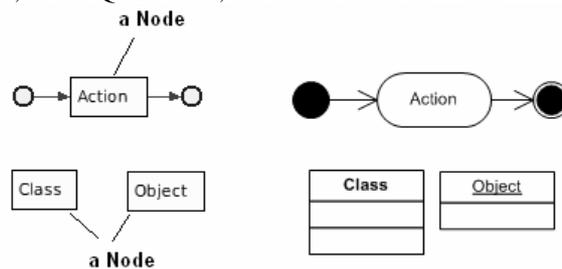

**Fig. 3. Topologos Nodes** - UML action, class and object (right) and their equivalent within Topologos (left). Formally distinguishing between these elements is not required (all are Nodes). Consequently, any element is an object, and potentially an activity or a class.

.

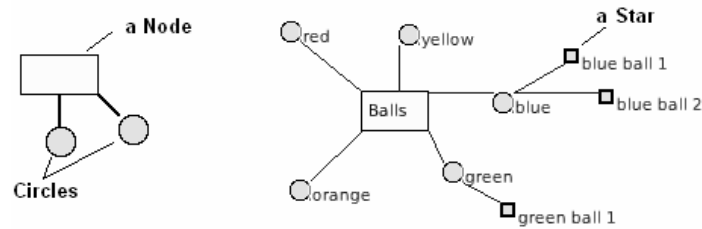

**Fig. 4. Topologos Circles** - Nodes may have several Circles containing objects. Circles are used for sorting the objects. Example: Node *Balls* has five Circles: *blue*, *green*, etc. Circle *blue* contains objects: *blue ball 1* and *blue ball 2* and Circle *green* contains *green ball 1*.

Figure 3 shows how Nodes may represent activities/actions of a UML Activity Diagram or objects or classes of an object model (UML Class Diagram).

The second part, called Circle, deals with the QOBJ places now dissociated from the node. Because we want to be able to sort the objects entering the activities, we may connect to a node as many circles containing objects as we wish (figure 4).

In order to allow an object to be simultaneously involved in one or more activities, the identity of the object is separated from the object itself. Topologos shows instantiated objects using a new element called Star (actually square boxes in the figures). A Node has as many Stars as it has different activities in which it takes part, or it has relations (links) with other objects.

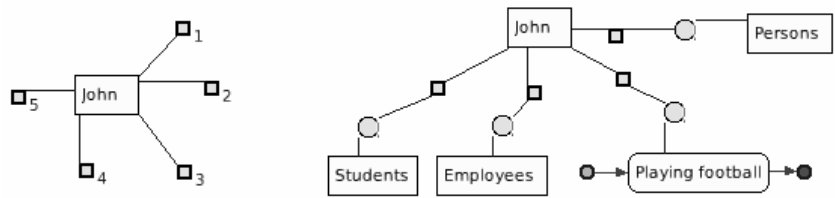

**Fig. 5. Topologos Stars** - (Left) John's identity is represented using Stars, shown as little square boxes in the figure: 1, 2, etc. (Right) Stars are used to represent John's identity within sets of objects (or classes), or processes in which he is involved: John is a Person, a Student, an Employee and is Playing football. Note that nodes get automatically rounded corners, if they appear to be part of a process, i.e. when arcs are connected to the nodes (see *Playing football*).

**Flows of objects and identity**

At this point, considering figure 5, we may want to express a conceptual difference between a person and a student or an employee. Who is John? He is a person. What is a Student? A person entering the process of studying. So we can transform the class *Students* into a process *Studying*.

Object flows are specified with a line starting from the Circle to the arc leading to the process as shown in figure 6.

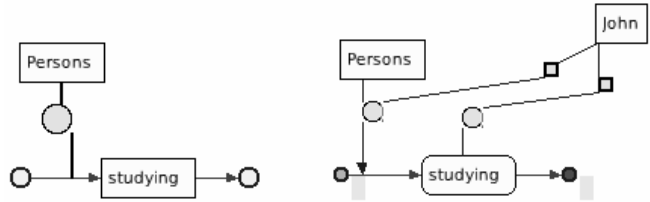

**Fig. 6. A student is a person in the process of studying** - Note the different representations of elements in figure 6 on the left side (without automatic layout) and the right side (with automatic layout). Topologos detects the type and meaning of the elements according to their topology. For instance start state and final states are automatically detected. Activities, i.e. Nodes within a process, have rounded corners. Relations between Circles and Arcs are transformed into arrows, indicating the flow of objects.

With the object-oriented approach every object has a unique identity. Modeling the categories of people of the university would require four classes: *Person*, *Student*, *Employee* and *Student-Employee* (for the case of students working at the university). If we want to model the *football players* as a separate category, we shall have three more classes: *Person playing football*, *Student playing football*, etc., which is quite prohibitive. Also, in order to allow an object to be simultaneously involved in one or more activities, the identity of the object (Star) is distinguished from the object, although this may be contradictory to the current object approach.

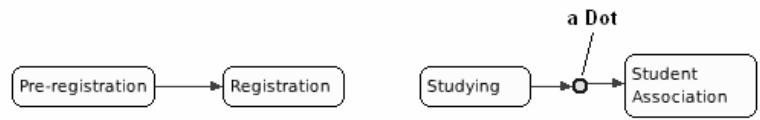

**Fig. 7. Topologos Dots** – (Rounded) Dots specify a duplication of identity. (Left) As within a usual UML Activity Diagram or Business Process Model, the objects leave the first stage *Pre-registration* while entering the next stage *Registration*. (Right) A *Dot* is inserted between the activity of *Studying* and the *Student Association*, because the related activies are not exclusive: members of *Student Association* are still *Studying*. Stars which represent the identity of the person are duplicated in both activities.

In a conventional simulation network, or UML Activity Diagram, the objects travel in the network from place to place, leaving the previous one while entering the next one. Because we can now distinguish between the object (Node) and its identity (Star), we may have transitions where identities are duplicated. We specify this behavior by placing a Dot in the transition (figure 7). A Dot is not only a pictorial element but also a sub-class of a Node. Hence it may have any characteristics or relations that a Node may have. Other Dot-like elements (not represented here) exist, such as square Dots for labelling object flows (see figure 16) or solid square Dots, also called Gate, which indicate the end of a token's life. For instance, at the end of study, the token of the person in the process of studying is destroyed. (A *Dot* refers per default to a rounded Dot).

**Pilots and services**

Important information within any process is to know who is in charge of the activity or action. Here we apply the concept of a pilot as described for the QOBJ. Pilots are specified using a root relation (i.e. a relation with a particular flag). It is possible to define a pilot for the whole process, and/or specify one as responsible for a particular action, as shown in figure 8.

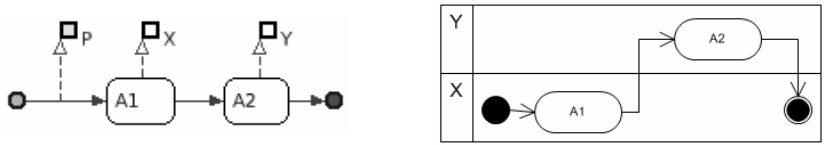

**Fig. 8. Topologos Pilots** – The figure shows how swimlanes of UML Activity Diagram (right) are replaced with Pilots (left). P is the owner of the whole process and responsible for all activities within the process. Control (and responsibility) of activity A1 (resp. A2) is delegated to pilot X (resp. Y). Pilots and services implement the concept of role, with the particularity that the identity of the role player is distinguished from his service (the service is the list of instructions that tells the pilot how to perform the control). Changing roles, i.e. the process control, is done by changing the pilot agent, or simply by changing his service.

**Association**

In order to combine the process model (UML Activity Diagram) and the class diagram (UML Class Diagram) we consider a class as a set of objects. Classes and Circles are both containers of objects. It is thereby possible to establish association relationships between sets of objects by connecting the circles of the process model as shown in figures 9 and 10, in a similar manner as with classes in a UML Class Diagram. Because Circles are not specific to object models, Association relationships may occur between an object model view and a process model view.

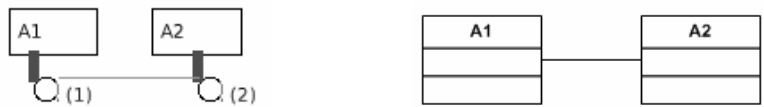

**Fig. 9. Association** - UML Class Diagram and class relationship (right), and its Topologos equivalent (left).

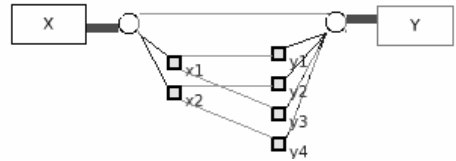

**Fig. 10. Relations between objects** - The relations between circles indicate that associations may be set between objects contained in the circles. The formal relations between circles are instantiated by the objects. Node X contains two objects: x1 and x2. The formal relation X – Y is instantiated between x1 (resp. x2) by objects y1, y3 (resp. y2, y4).

**Sub-sets and inheritance**

Because there are no conceptual differences between circles, it becomes possible to dynamically create new sets of objects, i.e. new classes. Since the association relationships between classes are explicitly laid out in the model, the notion of inheritance becomes useless. Figure 11 shows how Inheritance relationships are simply implemented with Topologos.

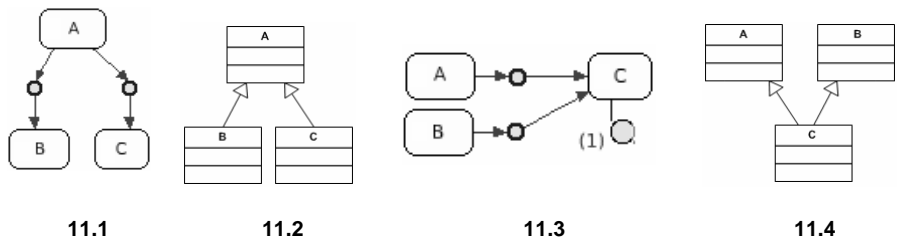

**Fig. 11. Sub-sets and inheritance** - Sub-classes within UML (11.2) are implemented as sub-sets within Topologos (11.1). Circle (1) in 11.3 contains objects which simultaneously belong to C and A (or B). Hence they have the properties of class A (or B) and C, which is characteristic of a multiple inheritance. Multiple inheritance is rarely used in the context of standard object-oriented modeling because of the complexity of managing the possible conflicts between attributes. These conflicts disappear with Topologos as the lexical information is replaced by a topological information. The position of the Circle in the network specifies the semantic of the attribute. Therefore the positions of the object's Stars (i.e. identities) accurately define the attributes' semantics.

**Merging Object Model and Process Diagram: an Example**

Figure 12 shows the process for organizing a meeting. The data structure related to each activity is represented in the diagram. For instance, activity 2: '*Select* *participants'* involves selecting participants from a list of persons, i.e. to establish a link between a *Meeting* object and the objects representing the participants at the meeting. The link between the *Meeting* and *Person* circles represents this relationship.

The participants are regrouped into the *Participant* circle. The dot (6) indicates that the participants are not moved from the *Person* circle, but their identity is duplicated in *Participant* circle.

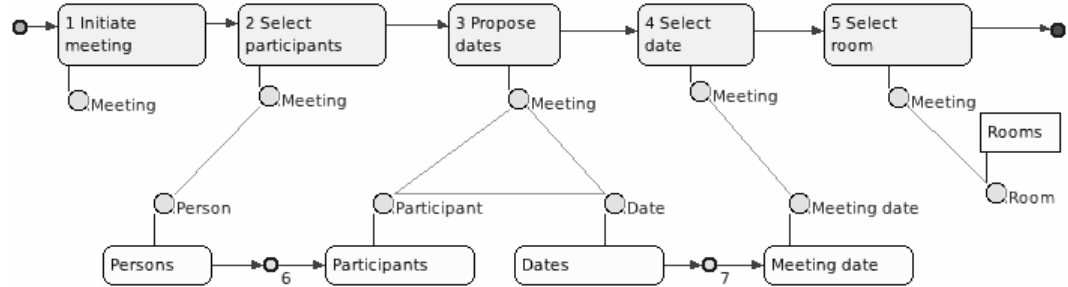

**Fig. 12. Meeting organization process** - The data structure related to each activity is represented in the diagram.

Figure 13 shows the meeting process with instantiated objects. To show how the object structure of an instantiated meeting object VIP is transformed by the process, we have laid out the five consecutive positions of the object in the same diagram.

**Fig. 13. Meeting organization process with instantiated objects (below)**

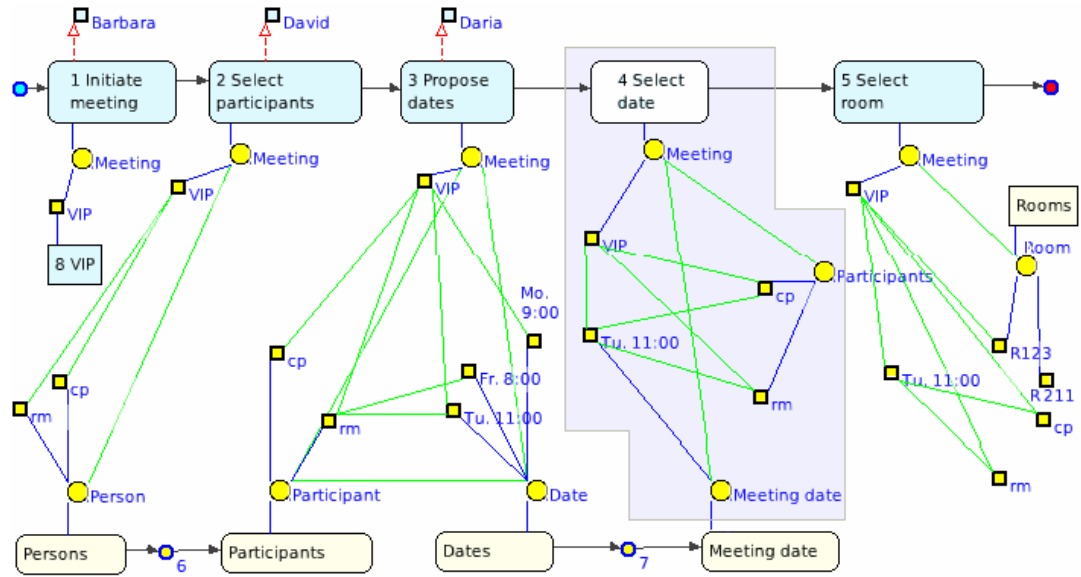

**Model usability and scalabity**

The example in figure 13 may seem complicated for such a simple framework as this, but some important points may be outlined:
- For the sake of explanation and for a lack of space in this paper, the five stages of one meeting are represented in figure 13. Only one stage is actually valid, in this instance stage 4, which has been highlighted.
- Both object and process models are shown in the figure. Topologos offers simple mechanisms to switch from one view to another, or to display both for a selected set of data. Showing the data in this way does not increase the complexity of the model view. The smooth transition between models makes it easier for the end-user to understand the relationship between their task (i.e. process view) and the data model than having separate object and process models.
- UML Class diagrams do not include (instantiated) objects, hence may look simpler than figure 13. Topologos hides the objects per default. It offers mechanisms to show only part of the network of objects. Allowing a selective monitoring of data, the user discovers only the view and the objects that concern him.
- Topologos avoids redundancy. To show that data produced by a process is used by another process (or part of a process), a simple relation between the corresponding circles can be set up.

The last features guarantee the scaleability of the model. A daily use of Topologos for project management and related documents shows how the *hide and show* mechanisms allow an amazing amount of information (i.e. objects: contacts, project stages and documents) to be shown in one model of terminal screen size. Moreover the number of models is unlimited.

To monitor a very large number of objects such as a complete list of students in a university it may be preferable to open separate windows showing conventional lists of objects. Topologos could be easily adapted in that manner. However, by using simple examples, the representation in figure 13 is sufficient to demonstrate to users what data is needed for the completion of their task and how they are transformed.

## 6  Discussion

Combining the process model with the class diagram (or object model) does not present any major difficulties. QOBJ Modeling is a good starting point for performing this amalgamation, but alternatively most process or data modeling tools could have been used. The main difficulty is conceptual. The decisive step consists of reconsidering the process of creating objects.

An object may now be created without any attributes, possessing only an identity. This identity gets dressed up with attributes (in relation to other objects) as the object travels through processes. As a result, the concept of inheritance seems to lose its significance. This conclusion could be difficult to accept. We would like to qualify it.

Firstly, the scope of our study is business modeling. As we previously stated, business data modeling should present each user only with the data related to his activity, and no superfluous data. In the scope of software development, the model obeys a software strategy, and it is often easier to introduce empty attributes than to create them dynamically.

Secondly, we must distinguish between two types of attributes: the ones which are defined at the creation of the object, and those which are dynamically added through later processes. The borderline between the two categories of attributes depends on the business domain. For instance, for a car manufacturer a car is the result of a manufacturing process, where different components are assembled. The car as an object does not have any attributes at the start. It has no wheel, no motor and no seat. A car, for a car rental agency, is already a completed object, with wheels, a motor, a color and a certain number of seats. New attributes may then be added, such as the customer name, date of contract, etc. The creation of the car object with its predefined attributes allows the hiding of creation processes outside of the business domain, i.e. hiding the processes of the environment. The classical approaches to object creation by instantiation of class, or object cloning, are not contradictory to the approach described here, but rather complementary.

**Process to integrate object and process models**

By automatically merging a process model and a class diagram, Topologos bridges the gap between object- and process- oriented models. Therefore, it is no longer necessary to decide beforehand which type of modeling will be used. This allows the person doing the modeling a greater flexibility. Information related to the data or the activities can be added to the model following discussion with an expert user.

After acquiring the user's knowledge, a phase of analyzing and consolidating the data may become necessary. There follows a rigorous process that allows integration of object and process models comprising the following steps:

- Step 1 : Establish a process model similar to BPM or UML Activity Diagram.
- Step 2 : Identify the tokens, i.e objects which cross the processes. The business information will be added to these tokens.
- Step 3 : Define the class of the tokens (and other objects) that will mask the creation process of data (object attributes) that is out of the scope of the business domain. Identify the static data of the token which could be inherited from a class (e.g. the name of a person). Ask yourself which attributes are pre-defined at the object creation and which ones should be dynamically created as the object crosses the processes.
- Step 4 : Identify the boundaries of identity (materialized with Dots : duplication of identity, or Gates: loss of identity).
- Step 5 : Define per activity (i.e. process stage) the necessary data to perform the task (inputs and intermediate data) and the data obtained at the end of the activity (data produced), set the relations to object Circles accordingly.
- Step 6 (option for monitoring): Finalize model and connect to target or productive system for monitoring.

## 7   A first pilot case study

To what extent do users accept the double (i.e. object and process) modeling approach? As an example we take a project performed at the University of Bern consisting of the installation of a system allowing the evaluation of the quality of teaching (education evaluation process). We will study it according to the method that we have defined in chapter 6.

**Step 1: Process model**

The principle means of the evaluating the quality of teaching involves distributing printed forms to the students. Once filled in, the forms are automatically inputted using scanning machines. Results are sent to the teachers and the University Headquarters. The whole process is under the responsibility of the Faculty. Some activities are delegated to the University Headquarters and to teachers.

The different stages of the process are shown in figure 14. The process ownership and delegation processes are described using the concept of Pilot. They will be added in the final model (see figure 16).

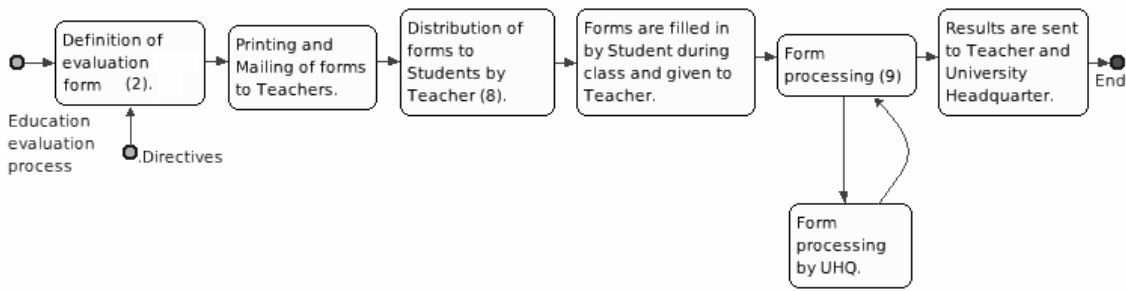

**Fig. 14. Education evaluation process** - The first activity (2) consists of designing the forms which will be given (in printed form) to the students. The forms' layout and contents follow the general *Directives* provided by the University Headquarters. The second activity involves printing the forms and adressing them to the teachers. Then the teachers carry out the third activity/task (8): they hand out the forms to the students during class. Forms are filled in by students and given back to their teacher. The processing of the forms occurs in the next stage (9). This may be undertaken by the Faculty or by the University Headquarters; therefore, the *Form processing* activity has been duplicated. Results of the evaluation are then sent to the teacher and the University Headquarters.

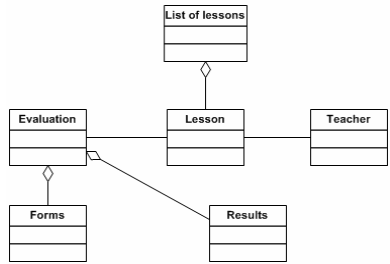

**Fig. 15. Data model for the Education evaluation process**

A conventional UML Class Diagram related to this process is shown in figure 15. This representation holds some ambiguities. For instance, the term *Form* could refer to the template of the form or to the forms once they have been filled in by the students. The aim of merging the process model in figure 14 and the data model in Figure 15 is to dispel such ambiguities, making the model more accurate and easier to understand for the users. The result is shown in figure 16.

**Step 2 Identify the token**

The process in figure 14 is to be followed every time a new evaluation is done. Hence the best and most natural token is the object *Evaluation*. Data related to the process will be added to this object at the stage when they are used or produced. The flow of tokens is indicated in figure 16 with the relation starting from Circle *Evaluation* (6) and connected to the first arc of the process (7).

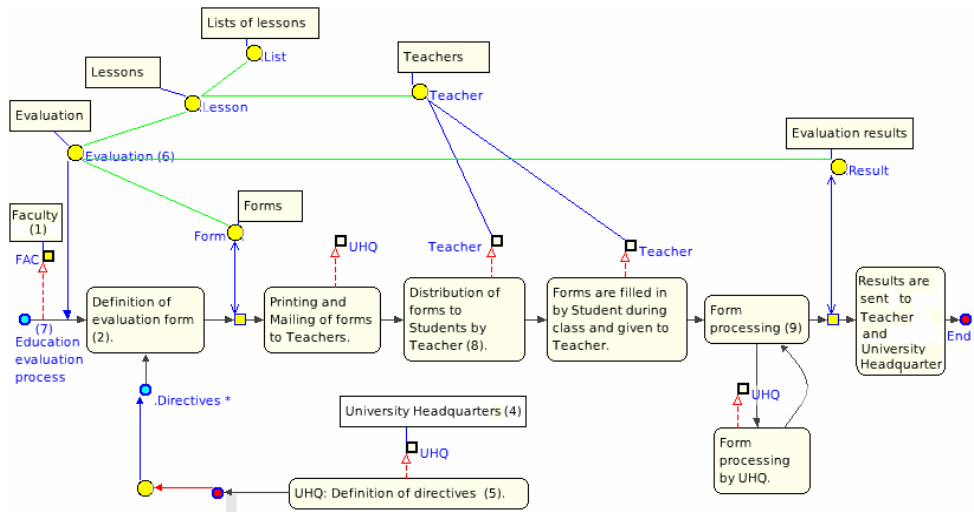

**Fig. 16. Merging Object and Process Diagrams for the Education Evaluation**

Figure 16 shows the process and the objects being processed, and indicates who is responsible for the whole process, i.e. in this case the *Faculty* (1) FAC, and to whom certain tasks have been delegated (for instance the *Distribution of forms* (8) is delegated to the *Teacher* declared as Pilot). Note that the activity of the students (filling in the Form) is an underlying process, independent of the computer system because forms are given to students as paper (printed) forms and remain anonymous; hence the students do not need to be modeled here. The *Evaluation* object (as token) moves through the process and thereby acquires new pieces of information (e.g *Evaluation form*, the result of stage (2), or E*valuation result*, product of stage (9)). Some external inputs may be necessary to perform activities, see for instance the *Directives* provided by the University Headquarters (4) resulting in an external process: *Definition of Directives* (5).

### Step 3 Static data

Static data, i.e. attributes which are already defined at the beginning of the process and which may stay invariant during the whole process, may be inherited using a convential class definition. For instance the *Lesson* which is evaluated could be set as attribute to the *Evaluation* object. But in doing this we may encounter difficulties establishing and showing the relationship between the *Lesson* and the *Teacher* (who plays a role in the process). Therefore we prefer to set this relationship explicitly using the Topologos approach (figure 16). This will moreover allow easy navigation of the model and easy monitoring of the process.

### Step 4 Boundaries of identity

Because the *Education evaluation process* does not have activities working in parallel, the *Evaluation* object crosses the whole process without duplication of identity. There is no need for Dots within this process. If the Forms were processed *both* by the Faculty and the University Headquarer, a Dot would be inserted between these two activities (see *Form Processing* (9) in figure 16).

Note that *Directives* enter as input into the process with a Dot (see *Definition of evaluation form* (2) in figure 16), obviously because they are not removed from the list of directives while being used in this process.

### Step 5 Identify data and object flows

The square boxes (square Dots) inserted between Nodes are used to label the object flows, i.e. the input/output of activities. Naming of an object flow is done with relation to the Circle that contains the object. Output of activity *Definition of evaluation form* (i.e. *Forms*, which are pre-requisite for the next activity), and result of activity *Form processing* (i.e. *Evaluation results)* are indicated in this way. Note that in the case of possible duplication of identity, the square Dots would simply be replaced by rounded Dots. The rounded or square Dots are not only pictural elements. They obey the basic concepts of Topologos. Because a Dot is a degenerated Node (implemented as a sub-class of Node), a relation between a Circle and a Dot indicates its contents.

With the previously shown examples (*Directives* input, *Evaluation* token), figure 16 includes the main ways to express *Objects Flows* using Topologos.

**User feedback**

The dual model as presented in figure 16 was well-accepted by the person in charge of the evaluation process and has been included in the project documentation as basis for the completion. We presume that the merging of process and object models led to a better understanding and acceptance because all elements of the evaluation process could be shown on the model with their relationships, including objects (forms, directives) and activities. However, we can not fully assess through this first pilot case study to what extent all the components of the model have been assimilated. Further studies should be conducted.

Topologos seems well-suited to various levels of modeling: from administrative processes (with many human interactions and information having a long lifespan) to fully automated and technical processes (services within a Service Oriented Architecture, or industrial real-time process control), as our experience with QOBJ modelling indicates, Topologos being an extension of this approach.

## 8  Conclusion

Even if Business Data Modeling and Software Data Modeling both use the Class Diagram as a tool, they are different activities which must be treated differently. Business information modeling is an instrument for dialogue between IT-experts and users, and consequently must deliver data models which are absolutely coherent to its users and consistent with their activities. This leads to the conclusion that a close bond must be established between Business Process Modeling and Business Data Modeling. Discrete event simulation tools manage flows of objects, and constitute an adequate basis for combining both models. Our development of a modeling tool has demonstrated the feasibility of this approach. However, as a consequence of this approach, the static object structure becomes dynamic. Further works will be necessary to refine the modeling paradigms and lead to a precise and fully adequate formalism.

We have proposed an approach to merging Process Diagrams and Class Diagrams, introducing Topologos as a modeling tool which enables their integration. We presented illustrated examples that demonstrate the limit of current modeling approaches based upon a dichotomy between process-oriented and object-oriented approaches, and the difficulty associated with modeling large scaled or dynamic data without taking into account their dynamic aspects, i.e. the processes which transform the data.

Certainly we can hardly expect to develop world-wide complex ecological models with the current modeling techniques having trouble modeling such simple cases as a student who becomes an employee of the university for just a month.

R. Descartes (1596-1650) wrote: « That the perceptions of the senses do not teach us the reality in things but only how these things are helpful or harmful to us. » [1] (our translation) [21]. This statement leads to questions as to the identity of things that we perceive, identify and finally try to model. In the frame of IT it is actually an illusion to think that we can shape an object whose forms, structures and relations with other objects would be set up, defined and remain unchanged independently of the user. The object is intrinsically multi-faceted, its identity depends on its utility and therefore on the person observing. Instead of talking about the identity of an object as a self, it would seem more valid to consider the identity of the object in association with a person and its utility to him. Thus the concept of identity holds a key position. The modeling approaches, especially the object-oriented modeling, have to consider instantiated objects and their identity as acquiring the same importance as processes and classes of objects.

The experimental tool Topologos comprises both 1/ the integration into a network of the object and process approaches, and 2/ the identity of the real instantiated objects.

The more models widen, the less IT modeling can elude such questions. With this contribution we hope to have engaged the discussion that actually is the main focus of this paper.

---

[1] « Nos sens ne nous enseignent pas la nature des choses, mais seulement en quoi elles nous sont utiles ou nuisibles. » R. Descartes, 1647.

## References

1. Alexandre Bergel, Stéphane Ducasse: "Scoped and Dynamic Aspects with Classboxes", L'objet – 9/2004.
2. Nathanael Schärli, Stéphane Ducasse, Oscar Nierstrasz, Andrew Black: "Traits: Composable Units of Behavior", Proceedings ECOOP 2003, LNCS, vol. 2743, Springer Verlag, pp. 248-274, July 2003.
3. Oscar Nierstrasz, Marcus Denker, Tudor Gîrba, Adrian Lienhard: "Analyzing, Capturing and Taming Software Change", www.iam.unibe.ch/~scg.
4. Nathanael Schärli, Stéphane Ducasse, Oscar Nierstrasz, Roel Wuyts: "Composable Encapsulation Policies", in Proceedings ECOOP 2004, LNCS 3086, pp. 248-274, Springer Verlag, 2004.
5. David Ungar, Randall B. Smith: "Self: The Power of Simplicity", originally published in OOPSLA '87 Conference Proceedings (SIGPLAN Notices, 22, 12 (1987)227-241).
6. Patrick Chénais: "Method and apparatus for assigning calls entered at floors to cars of a group of elevators", Schindler US Patent 5612519, 1997.
7. Patrick Chénais: "Safety equipment for multimobile elevator groups", Schindler US Patent 5877462, 1999.
8. Antonio Stagno, Patrick Chénais: "QOBJ: A Unification of Objects and Queuing Networks", Mathematical Modeling and Scientific Computing, Vol. 6, Boston 1996.
9. Antonio Stagno, Patrick Chénais, Thomas M. Liebling, Rémy Glardon: "QOBJ modeling or when existing simulation concepts do not apply well", Proceedings of 1st World Congress on Systems Simulation (WCSS97), Singapour 1997.
10. Antonio Stagno, Patrick Chénais, Thomas M. Liebling: "QOBJ Modeling - A new Approach in Discrete Event Simulation", OR Spektrum 20 (1998)109, 1998.
11. Paul T. Ward: "How to Integrate Object Orientation with Structured Analysis and Design", IEEE Software, March 1989.
12. Xavier Briffault, Stéphane Ducasse, Squeak Programmation, Collection Coming next, Eyrolles, 2001.
13. Jian Zhong: "Step into the J2EE architecture and process, Develop complete J2EE solutions with an eight-step cycle", JavaWorld.com, September 2001.
14. Martin Fowler, Kendall Scott: UML Distilled. Addison-Wesley Longman, 1997.
15. Craig Larman: Applying UML and Patterns, Prentice Hall Inc., 2002.
16. Donald Bell: "UML basics. Part III: The class diagram", *IBM Global Services.*
17. Grady Booch, James Rumbaugh, Ivar Jacobson: The Unified Modeling Language User Guide. Addison Wesley Longman, 1999.
18. James Rumbaugh, Ivar Jacobson, Grady Booch: UML 2.0 Guide de référence. Campus Press, 2004.
19. Dan Pilone, Neil Pitman: UML 2.0 in a Nutshell, O'Reilly, 2005.
20. Philippe Desfray, A method for object oriented programming: The class-relationship method, SOFTEAM. Refer to Object engineering - the fourth dimension - Philippe Desfray Addison-Wesley, 1994. ISBN 0-201-42288-3).
21. René Descartes, Principes de la philosophie 1647.
22. Tom DeMarco: Structured analysis and system specification. Englewood Cliffs - N.J. Prentice-Hall, 1978.
23. Patrick Chénais, Karl Weinberger: "New Approach in. the Development of Elevator Group Control. Algorithms", ELEVCON, Amsterdam 1992.

## About the author

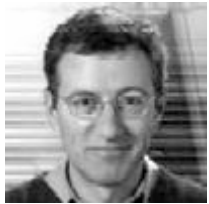

Patrick Chénais received his doctorate title in Computer Science in March 1987 from the University of Toulouse. Then he joined the R&D center of Schindler Elevator, in Ebikon/Lucerne, Switzerland. Afterwards he was responsible for the development of the Schalter Automation Post Project which involved the automation of Swiss post office counters. Later he was for five years in charge of the development of administrative data processing at the University of Bern, Switzerland. He is now working at the Federal Office of Information Technology, Systems and Telecommunication FOITT in the eGovernment department and can be reached at patrick.chenais[at]bit.admin.ch.